\pdfoutput=1 
\documentclass[times]{cnmauth}

\usepackage{moreverb}

\usepackage[colorlinks,bookmarksopen,bookmarksnumbered,citecolor=red,urlcolor=red]{hyperref}

\usepackage{url,lineno,microtype,subcaption}
\usepackage[onehalfspacing]{setspace}
\usepackage{color} 

\usepackage{import}
\usepackage{epsfig}

\usepackage{paralist}

\usepackage{amssymb}
\usepackage{amsmath}
\usepackage{bm}

\usepackage[mathscr]{euscript}
\usepackage{comment}

\usepackage[noabbrev,capitalise,english]{cleveref} 

\usepackage{framed} 
\usepackage{multicol} 

\usepackage{multirow} 

\usepackage[noprefix]{nomencl} 
\makenomenclature
\graphicspath{{./figs/}}

\usepackage{xcolor}
\colorlet{lightred}{red!70!white}

\usepackage[noadjust]{cite}

\usepackage{blindtext}

\newcommand{\eg}{e.g.} 
\newcommand{\ie}{i.e.} 

\makeatletter
\newcommand*{\rom}[1]{\expandafter\@slowromancap\romannumeral #1@}
\makeatother

\usepackage{lineno}

\newcommand\BibTeX{{\rmfamily B\kern-.05em \textsc{i\kern-.025em b}\kern-.08em
T\kern-.1667em\lower.7ex\hbox{E}\kern-.125emX}}

\def\volumeyear{2017}

\begin{document}

\runningheads{H.~P.~Bui et al.}{Controlling the Error on Target Motion through Mesh Adaptation}


\title{Controlling the Error on Target Motion through Real-time Mesh Adaptation: Applications to Deep Brain Stimulation}

\author{Huu Phuoc Bui\corrauth \,$^{1}$, Satyendra Tomar\,$^{1}$, Hadrien Courtecuisse\,$^{2}$, Michel Audette\,$^{3}$, St\'{e}phane Cotin\,$^{4}$ and St\'{e}phane P.A. Bordas\corrauth \,$^{1,5,6}$}

\address{
\begin{center}
 $^{1}$Institute of Computational Engineering, University of Luxembourg, Faculty of Sciences Communication and Technology, Luxembourg \\
$^{2}$University of Strasbourg, CNRS, ICube, Strasbourg, France \\
$^{3}$Department of Modeling, Simulation and Visualization Engineering, Old Dominion University, Norfolk, USA \\
$^{4}$Inria Nancy-Grand Est, Villers-l\`{e}s-Nancy, France \\
$^{5}$Institute of Mechanics and Advanced Materials, School of Engineering, Cardiff University, UK \\
$^{6}$Intelligent Systems for Medicine Laboratory, University of Western Australia, Perth, Australia
\end{center}
}

\corraddr{huu-phuoc.bui@alumni.unistra.fr (H. P. Bui), stephane.bordas@alum.northwestern.edu (S. P. A. Bordas)}

\begin{abstract}

We present an error-controlled mesh refinement procedure for needle insertion simulation and apply it to the simulation of electrode implantation for deep brain stimulation, including brain shift.

Our approach enables to control the error in the computation of the displacement and stress fields around the needle tip and needle shaft by suitably refining the mesh, whilst maintaining a coarser mesh in other parts of the domain.

We demonstrate through academic and practical examples that our approach increases the accuracy of the displacement and stress fields around the needle without increasing the computational expense. This enables real-time simulations.

The proposed methodology has direct implications to increase the accuracy and control the computational expense of the simulation of percutaneous procedures such as biopsy, brachytherapy, regional anesthesia, or cryotherapy and can be essential to the development of robotic guidance.

\end{abstract}

\maketitle


\section{Introduction}

Real-time simulations are increasingly frequent for many applications, from physically based animation \cite{nealen2006physically,wang2015linear} to medical simulation \cite{Courtecuisse2014}. Within the medical context, simulations involving interactions between an interventional radiologist or a surgeon with soft, deformable organs, are helpful. Indeed, these simulations do not only have the potential to help surgeons train or plan  complex operations but they can also guide them during the intervention itself. In spite of the obvious importance of quality control in such safety-critical advanced numerical tools, no approach is known to evaluate and control the error committed by simulations. If this were possible, not only would the user be provided with useful information regarding the quality of the results, but the computational cost could be focused on where the error is large and decreased where it is small, thus leading to computational savings.

Yet, whilst the literature in error estimation is rich within the computational mechanics community, little or no work has been done in the context of medical simulation, in particular for real-time applications \cite{Wu2001}. The error in numerical simulation comes in three guises. First, the user writes a mathematical model of the system of interest. For example, they may choose a corotational model to represent the tissue, or a hyper-elastic model with a number of parameters. The inability of the model to represent the physical reality is measured by the \emph{model error}.

Second, the mathematical model must be solved numerically, since analytical solutions are rarely available. For example, a user might wish to use the finite element method (FEM), or meshfree methods, or the boundary element method. The inability of a numerical method to exactly solve the mathematical model leads to the \emph{discretisation error}.

Third, the set of linear equations provided by the discretised model must be solved numerically. The associated error is known as the \emph{numerical error} and incorporates, \eg{} round-off errors.

In this paper, we tackle only discretisation error. Our assumption is that the mathematical model used (here the corotational formulation), properly represents the behaviour of the organ in question, in our case the brain. This is a strong assumption which is most likely not verified in general, as it was shown that brain matter behaves hyper-visco-elastically \cite{Mihai2015}. 

Yet, the tools we develop in this paper are directly applicable to any model which can be described in the form of partial differential equations, which is the case of most models of soft tissues. The aim of this paper is therefore to demonstrate that error estimation techniques can be effectively used, even in real-time scenario, to control the discretisation error and drive local refinement of the discretisation. We focus on the problem of needle insertion because of the particular challenges it poses, due to the localisation of deformations and stresses close to the needle shaft.

In modern clinical interventions such as biopsy, brachytherapy, cryotherapy, regional anesthesia, or drug delivery, needle-based percutaneous procedures play an important role, thanks to their minimally invasive character. Good training and careful planning to optimise the path to the target, while avoiding critical internal structures, directly influence the desired achievement of these procedures \cite{Hamze2016}. In some instances, robotic devices can be used to assist these procedures~\cite{Elgezua2013}. Unfortunately, natural tissue motion before or during the intervention, and organ deformation due to interactions with the needle during insertion, generally lead to inefficient or incorrect planning~\cite{Hamze2016}. An accurate simulation of needle insertion can address these issues, possibly in a manner complemented by imaging in intraoperative scenarios. In addition to the simulation accuracy, the computational speed is vital, for most problems, since the biomechanical simulation is a kernel part of an optimised algorithm for the needle trajectory or for a robotic control loop.

In needle insertion simulations, one has to deal with three main models: a model for the soft tissue, a model for the flexible needle, and a model described the needle-tissue interaction. Needle and tissue models can rely on linear or non-linear constitutive laws (see the survey in \cite{Abolhassani07}). A nonlinear viscoelastic Kelvin model is used for the needle in \cite{Mahvash2009}. The model describing the interaction between the needle and the tissue remains a main challenge. It consists of a number of different physical aspects, such as frictional contact on tissue surface, puncturing, cutting, sliding with friction, and the Poynting's effect. Dehghan et al. \cite{Dehghan08} propose an ultrasound-based motion estimation to model needle-tissue interaction. To avoid obstacles while steering the needle during insertion, different kind of needles have been proposed, for example, a symmetric tip \cite{Heverly2005}, a flexible bevel tip \cite{Alterovitz05,Webster06,Misra2010}, as well as a compliant needle model \cite{Tang08}.

FEM is the most commonly used numerical method for needle insertion simulations, see~\eg{} \cite{DiMaio03,Hing2006,Jiang2008}. In order to numerically solve the equilibrium equations of mathematical models, FEM uses the concept of discretisation, in which the volume of an object/organ is decomposed into elements (such as tetrahedra or hexahedra). The arrangement of these elements defines the mesh, and the continuous mathematical problem, in this case biomechanical deformation, can be established on this mesh. The resulting discrete problem (refer to \cref{sec:weakForm}), with the number of unknowns directly proportional to the number of elements in the mesh, can then be easily solved on a computer. It is noted that the computational cost (time) is also directly proportional (with cubic power) to the number of unknowns, \ie{}, the CPU time is of the order of $N^3$, where $N$ denotes the number of unknowns of the problem \cite{Watkins2005}.

Using a uniform coarse mesh (involving few elements of the same shape) is undermined by the limitation that discontinuities engendered by cuts, singularities or stress concentrations can not be captured. On the other hand, it is uneconomical, or in a strict sense, prohibitively expensive, to use a uniformly fine mesh (involving same shaped elements but many more as compared to a coarse mesh). Therefore, in the context of real-time simulations, an economical hybrid mesh should be used to deliver fast simulation response while still enabling accurate simulation results.

Our aim is to model the interactions between the needle and the tissue employing an adaptive meshing technique, which is guided by an a posteriori error estimate. An a posteriori error estimation approach obtains the actual error bounds using the computed solution (instead of the asymptotic bounds on the solution of the original mathematical problem). In this paper we use an a posteriori error estimation technique called super-convergent patch recovery (SPR). The SPR technique was proposed by Zienkiewicz and Zhu \cite{zienkiewicz1992superconvergent}, and the asymptotic convergence of this technique is studied in \cite{CarstensenB-02-Apost,BartelsC-02-Apost}.
By solving the discrete problem on a computer, we obtain the numerical solution (henceforth called the \emph{raw} solution as in \cite{Bordas2008}). In the SPR (see \cref{sec:SPR} for details), we post-process the \emph{raw} solution to obtain an \emph{improved} solution, to which the \emph{raw} solution is compared.
The key idea is that, in certain regions, where the approximated error (\ie{} the difference between the \emph{raw} solution and the \emph{improved} solution) is greater than a certain threshold, the mesh should be refined locally there, and where the error is small (these two solutions are close together), the mesh can be kept unchanged or coarsened. Note that the mesh refinement is only introduced as a measure to improve the simulation accuracy of the interactions between the needle and the tissue. Similar to the approach used in \cite{Duriez2009}, we do not require the mesh used for the tissue to conform to the needle path.

The efficiency of our method will be studied through a heuristic needle insertion scenario, and also through a more complex simulation of the insertion of an electrode lead which is used in Deep Brain Stimulation (DBS). The latter is a surgical procedure being able to treat a number of disabling neurological symptoms, \eg{} Parkinson's disease (PD). DBS consists of inserting a long needle-shaped electrode (or electrode lead) through a small opening in the skull into the brain to reach the target area which for PD is the subthalamic nucleus (STN). Subsequently, a neurostimulator sends therapeutic electrical impulses to this target implanted with electrodes. The success of the therapy largely depends on the exact implantation of the electrode lead \cite{Breit2004}, although increasingly, DBS targeting benefits from knowledge of afferent and efferent white matter tracts \cite{Calabrese2016}. This procedure necessitates a pre-operative planning step for determining the target coordinates to implant the electrodes, as well as a trajectory to deliver the insertion cannula, through which then one determines the electrode insertion path. A pre-operative magnetic resonance imaging (MRI) of the patient is employed to identify and locate the target within the brain. However, during the DBS procedure, as a burr hole is drilled in the skull to access the brain tissue, brain shift occurs due to the leak of cerebro-spinal fluid \cite{Bilger2011,Hamze2015_brainShift}. This includes an error in the planning, that can at best make the therapy ineffective, and at worst cause complications, such as psychiatric complications \cite{piasecki2004psychiatric}. Therefore, brain shift prior to electrode insertion will be taken into account in our simulations.

The remaining of the paper is organised as follows. In \cref{sec:Method}, we describe the problem being solved. The interaction model between needle and tissue is formulated, and it is described how the elements are marked for refinement during simulation by employing an error estimate. It is then followed by the discussion on solving the system equations with constraints. Numerical results are presented in \cref{sec:Results}, which shows the efficiency of the proposed approach through a simple needle insertion scenario, as well as through more complex DBS lead insertion simulation. Finally, conclusions are drawn in~\cref{sec:Conclusions}.

\section{Methods}
\label{sec:Method}

We first outline the mathematical model we choose to represent the brain and its interaction with the cannula/needle/lead. We then describe the approach we follow to solve this problem numerically.

We use corotational elasticity, which implicitly assumes that the deformations of the brain are small. We develop a frictional needle-tissue interaction model. The problem is solved using hexahedral finite elements implemented in the open-source Simulation Open Framework Architecture (SOFA, \url{www.sofa-framework.org}). The local $h$-refinement strategy is based upon simple SPR-based a posteriori error estimation method.

\subsection{Problem statement}
\begin{figure}[!htbp]
 \centering
\includegraphics[width=0.6\columnwidth]{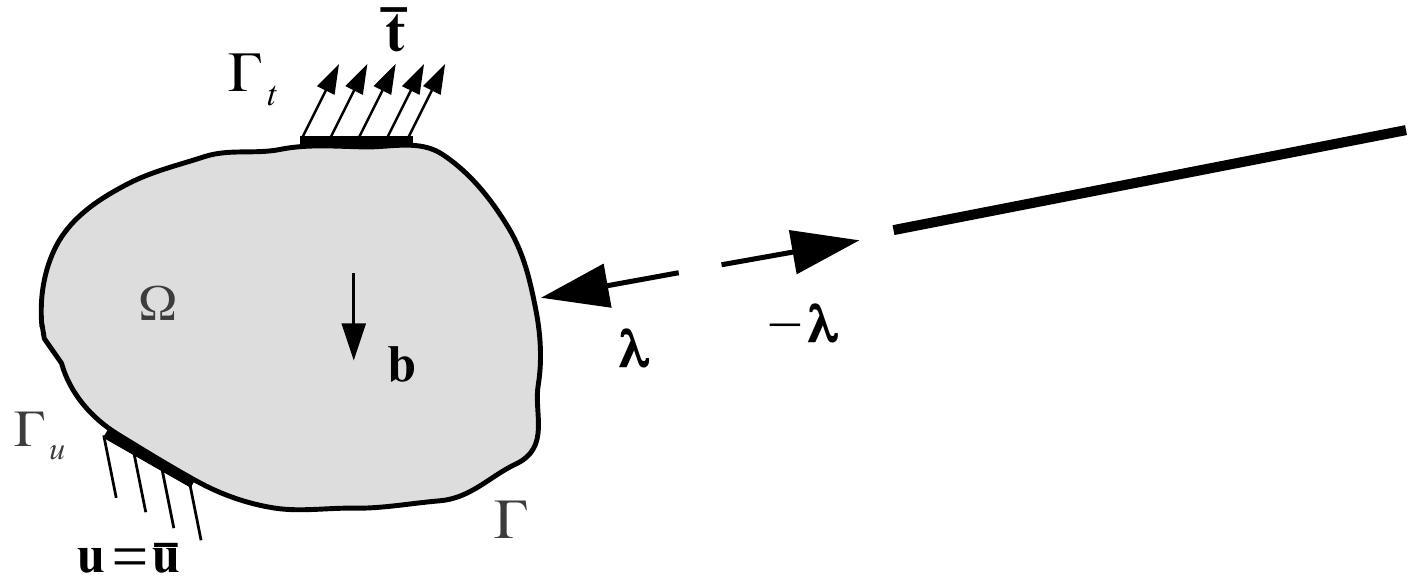}
\caption{An object $\Omega$ undergone a traction force $\mathbf{\bar{t}}$, a body force $\mathbf{b}$, a prescribed displacement $\mathbf{\bar{u}}$, and is in contact with another object through interaction forces $\bm{\lambda}$.}
\label{fig:fem_tissue_needle}
\end{figure}

Within the context of needle (or cannula/lead) insertion into soft tissue, both the needle and the tissue are modeled as dynamic deformable objects. Let $\Omega$ represent the volume of an object (\eg{} the tissue). Undergoing external force $\bar{t}$, which is applied on the boundary part $\Gamma_t$ on the object boundary $\Gamma$, and bearing a prescribed displacement $\mathbf{\bar{u}}$ on its boundary part $\Gamma_u$ in $\Gamma$, see~\cref{fig:fem_tissue_needle}, the dynamic equilibrium motion equation of the object is expressed by, see~\cite{zienkiewicz2000finite,Liu201443},
\begin{equation}
 \mathrm{div}\, \bm{\sigma} + \mathbf{b} + \bm{\lambda} =  \rho \ddot{\mathbf{u}} \qquad \text{in } \Omega .
 \label{eq:equilibrium}
\end{equation}
Here $\bm{\sigma}$ is the Cauchy stress tensor (accounting for internal forces), $\mathbf{b}$ is the body force vector (\eg{} gravity), $\bm{\lambda}$ is the interaction force with another object (\eg{} a needle), $\rho$ denotes the mass density, $\mathbf{u}$ stands for the displacement field of the object, and $\ddot{x}$ is the second partial derivative of $x$ with respect to time. In physical terms, the divergence term ($\mathrm{div}\, \bm{\sigma}$) is the sum of the derivatives of the stress tensor components with respect to each axis direction. The compatibility conditions on the boundaries $\Gamma_t$ and $\Gamma_u$ read
\begin{subequations}
\begin{align}
  \bm{\sigma} \cdot \mathbf{n} & = \mathbf{\bar{t}} \qquad \text{on } \Gamma_t \label{eq:NeumannBC}\\
 \mathbf{u} & = \mathbf{\bar{u}} \qquad \text{on } \Gamma_u. \label{eq:DirichletBC}
 \end{align}
 \label{eq:compatibility}
\end{subequations}
\cref{eq:NeumannBC,eq:DirichletBC} are known as Neumann and Dirichlet boundary conditions, respectively.

The strain in the object, which is a measure of the object deformation, can be expressed from the gradient of the displacement as
\begin{equation}
  \bm{\epsilon}  =  \frac{1}{2}\left(\mathrm{grad}\,\mathbf{u} + (\mathrm{grad}\,\mathbf{u})^T \right).
  \label{eq:kinematic}
\end{equation}
The constitutive law, that expresses the relation between stress and strain tensors (through a function $f$) of the object via the model internal (intrinsic) variables $\bm{\nu} = (\nu_1, \nu_2, \dots, \nu_n)$, reads
\begin{equation}
  \bm{\sigma} =  f(\bm{\epsilon},\bm{\nu}).
  \label{eq:constitutive}
\end{equation}
\cref{eq:equilibrium,eq:compatibility,eq:kinematic,eq:constitutive} constitute the governing equations that describe the behaviour of the object in the sense of continuum mechanics.

\begin{figure}[!htbp]
\centering
\includegraphics[width=0.35\columnwidth]{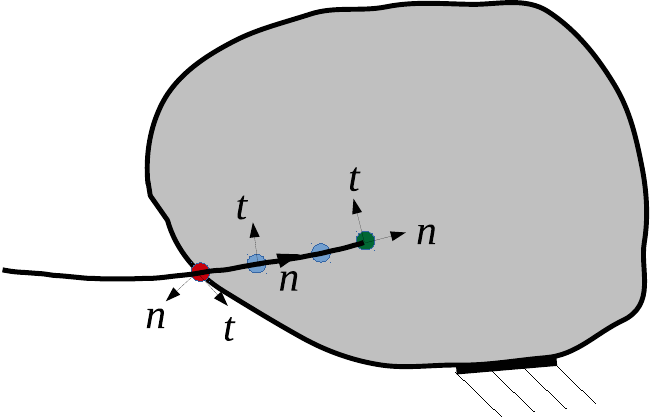}
\caption{As in \cite{Bui2016_TBME}, three types of constraints between the needle and soft tissue are defined: surface puncture (denoted by $\bm{\lambda}^{ts}$, illustrated in red), needle tip constraint (denoted by $\bm{\lambda}^{nt}$, illustrated in green) and needle shaft constraints (denoted by $\bm{\lambda}^{ns}$, illustrated in blue). To express the constraint directions, we define, at each constraint point, a local coordinate system $n$-$t$.}
\label{fig:needle_tissue_schema}
\end{figure}

The interaction force $\bm{\lambda}$ is defined from the law describing the needle-tissue interaction. During the needle insertion, we prescribe three different types of constraints between the needle and the tissue as in \cite{Bui2016_TBME}, see \cref{fig:needle_tissue_schema}.
The Coulomb's friction law is employed to describe the frictional contact within these three types of constraints. First, we define a puncture constraint between the needle tip and the tissue surface. This constraint follows the Kuhn-Tucker conditions. The latter describes the fact that the contact force only exists when the needle tip is in contact with the tissue surface. In the direction $n$ (normal to the tissue surface), this condition reads
\begin{equation}
\delta_n \geq 0, \qquad \lambda_n^{ts} \geq 0, \qquad \delta_n \cdot \lambda_n^{ts} = 0,
\end{equation}
where, the superscript $ts$ stands for the constraints on the tissue surface, $\delta_n$ is the distance between the needle tip and the tissue surface in the direction $n$, and $\lambda_n^{ts}$ is the contact force in the normal (to the tissue surface) direction. Let $\lambda_{p_{0}}$ express the puncture strength of the tissue. When the interaction force is greater than the puncture strength of the tissue
\begin{equation}
\lambda_n^{ts} > \lambda_{p_{0}},
\end{equation}
the needle can penetrate into the tissue. And while the needle tip is in contact with the tissue surface but has not penetrated into it yet, the relative motion between them can be described by a frictional law. The Coulomb's friction law is employed to take into account the stick\slash slip between the needle tip and the tissue surface
\begin{equation}
\lambda_t^{ts} < \mu \lambda_n^{ts}  \quad \textrm{(stick)}; \quad \lambda_t^{ts} = \mu \lambda_n^{ts}  \quad \textrm{(slip)},
\end{equation}
in which the subscript ${t}$ denotes the tangent component of the contact force, $\mu$ is the friction coefficient.

Second, as soon as the needle penetrates into the tissue, we define a constraint between the needle tip and the tissue. The fact that the needle tip is stuck or can cut and advance in the tissue depends on the relationship between the normal component (in the direction $n$, along the needle shaft) and the tangent component (in the direction $t$) of the contact forces (see \cref{fig:needle_tissue_schema}), defined by
\begin{equation}
\lambda_n^{nt} < \mu \lambda_t^{nt} + \lambda_{c_{0}} \; \textrm{(stick)}; \quad \lambda_n^{nt} \geq  \mu \lambda_t^{nt} + \lambda_{c_{0}}  \; \textrm{(cut and slip),}
\end{equation}
where, the superscript $nt$ stands for the constraints at the needle tip, and $\lambda_{c_{0}}$ denotes the cutting strength of the tissue.

Finally, we define the needle shaft constraints (denoted by the superscript $ns$) along the needle shaft. These constraints help to enforce the needle shaft to follow the insertion trajectory created by the advancing needle tip. Again, we employ the Coulomb's friction law to describe the stick and slide contacts between the tissue and the needle shaft, expressed as
\begin{equation}
\lambda_n^{ns}  < \mu \lambda_t^{ns} \; \textrm{(stick)}; \quad \lambda_n^{ns}  = \mu \lambda_t^{ns} \; \textrm{(slide).}
\end{equation}

Note that these three types of constraints describe different kinds of physical interactions between the needle and the tissue. However, we depict three of them as interaction constraint vector, collectively denoted by $\bm{\lambda}$ in the following, which can be solved numerically.

\subsection{Weak form}
\label{sec:weakForm}

\begin{figure}[!htbp]
\centering
\includegraphics[width=0.6\columnwidth]{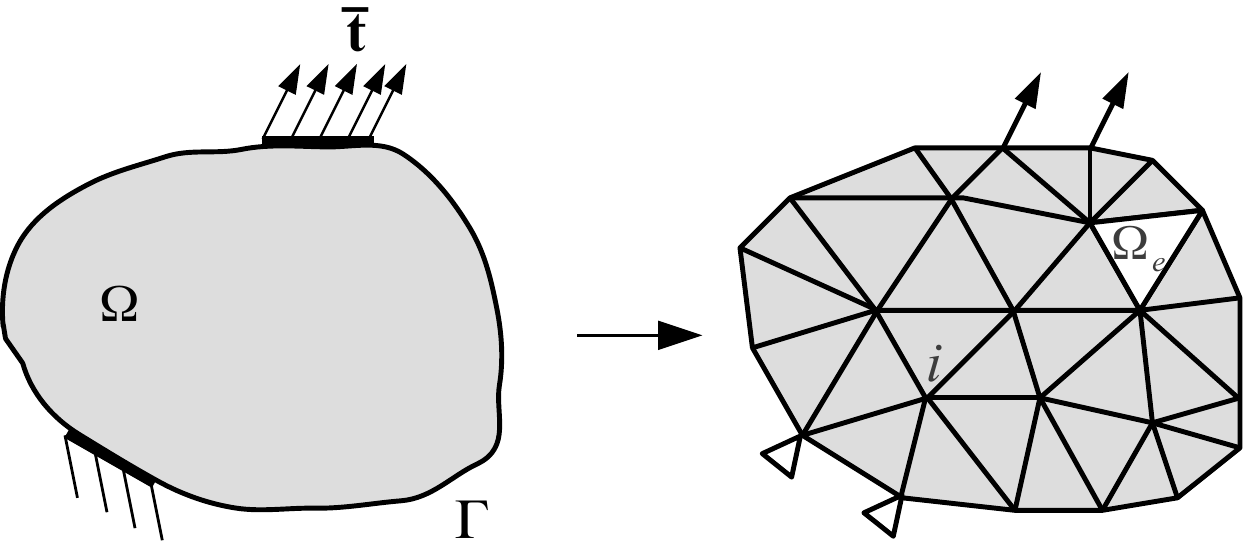}
\caption{Simplified illustration of FEM discretisation in two dimensions.}
\label{fig:fem_discretisation_schema}
\end{figure}

Since the partial differential equation \eqref{eq:equilibrium}, which states the equilibrium of the system, involves both spatial and temporal derivatives, it can be solved numerically by discretising that equation in both space (the volume representing the object) and time.

Using $N_n$ nodes, the domain $\Omega$, which represents tissue or needle, is spatially discretised into $N_e$ finite elements $\Omega_e$, $e=1,2,\dots,N_e$, see Figure~\ref{fig:fem_discretisation_schema}. By integrating the equilibrium equation on each element volume, and assembling for the whole volume, we obtain the discrete problem as (see,~\eg{},~\cite{Liu201443,zienkiewicz2000finite})
\begin{equation}
\label{eq:weakForm}
 \mathbf{M} \ddot{ \mathbf{u} } + \mathbf{C} \dot{ \mathbf{u} } + \mathbf{K} \mathbf{u} = \mathbf{f}^{ext} + \mathbf{H}^T \bm{\lambda},
\end{equation}
with $\mathbf{M}$ the mass matrix, $\mathbf{K}$ the stiffness matrix, $\mathbf{C}$ the damping matrix, and $\mathbf{f}^{ext}$ the external force vector. The interaction force constraints $\bm{\lambda}$, between the needle and the tissue, are computed using Lagrange multipliers (in numerical methods, the constraints $\bm{\lambda}$ are called Lagrange multipliers, and thus we use the same notation), where $\mathbf{H}^T$ provides the direction of the constraints.
\cref{eq:weakForm} can be rewritten as
\begin{equation}
\label{eq:Ma=f}
 \mathbf{M} \mathbf{a} = \mathbf{f}(\mathbf{x},\mathbf{v}) + \mathbf{H}^T \bm{\lambda},
\end{equation}
where $\mathbf{a} = \ddot{\mathbf{u}}$, $\mathbf{v} = \mathbf{\dot{\mathbf{u}}}$, $\mathbf{x}$ are the acceleration, velocity and position vectors, respectively, and $\mathbf{f}(\mathbf{x},\mathbf{v}) = \mathbf{f}^{ext}-\mathbf{K}\mathbf{u} - \mathbf{C}\mathbf{v}$ is the net force (the difference between the external and internal forces) applied to the object. Note that the displacement vector $\mathbf{u}$ is expressed through the current and initial position vectors, $\mathbf{x}$ and $\mathbf{x}_0$, respectively, as $\mathbf{u} = \mathbf{x} - \mathbf{x}_0$.
%

%
For temporal discretisation, i.e. to numerically solve the problem in time, we use an implicit backward Euler scheme~\cite{Baraff1998}, which is described as follows
\begin{equation}
\label{eq:EulerBackward}
\ddot{\mathbf{u}}_{t+\tau} = \frac{\dot{\mathbf{u}}_{t+\tau} - \dot{\mathbf{u}}_t }{\tau} ; \quad  \dot{\mathbf{u}}_{t+\tau} = \frac{\mathbf{u}_{t+\tau} - \mathbf{u}_{t}}{\tau},
\end{equation}
where $\tau$ stands for the time step. Inserting \cref{eq:EulerBackward} into \cref{eq:Ma=f} yields the final discrete system
\begin{equation}
\label{eq:Ax=b}
\underbrace{(\mathbf{M}-\tau \mathbf{C}-\tau^2 \mathbf{K})}_{\mathbf{A}} d\mathbf{v} = \underbrace{ \tau\mathbf{f}(\mathbf{x}^t,\mathbf{v}^t) + \tau^2 \mathbf{K}\mathbf{v}^t}_{\mathbf{b}} + \mathbf{H}^T \bm{\lambda}
\end{equation}
or simply $\mathbf{A} d\mathbf{v} = \mathbf{b} + \mathbf{H}^T \bm{\lambda}$, where $d\mathbf{v} = \mathbf{v}_{t+\tau} - \mathbf{v}_t$.
After solving~\eqref{eq:Ax=b} for $d\mathbf{v}$, the position and velocity are updated for needle and tissue as
\begin{equation}
\label{eq:updateVX}
\mathbf{v}_{t+\tau} = d\mathbf{v} + \mathbf{v}_t; \quad \mathbf{x}_{t+\tau} = \mathbf{x}_{t} + \tau \mathbf{v}_{t+\tau}.
\end{equation}

In this contribution, a lumped mass matrix, in which a diagonal mass matrix (from the mass density $\rho$) is integrated over the volume of each element is employed. The stiffness matrix $\mathbf{K}$ is computed based on the corotational FEM, in which the rigid body motion from total finite element displacements is extracted (since it does not contribute to element deformations, see~\cite{Felippa2005}). Note that the corotational FEM formulation makes it possible to handle large rotations for both, needle as well as tissue. For higher accuracy of the computed strain field and less sensitivity to locking phenomena, the soft tissue domain is discretised using hexahedral elements having $3$ translational degrees of freedom per node\footnote{It is also possible to use smoothed tetrahedral elements, which do not lock, see, \eg{} \cite{Nguyen-Xuan2010,Lee2017,Mendizabal2017}.}. We use a mesh that does not conform to the boundary of the domain, similar to  the approach of Immersed Boundary Method~\cite{Pinelli2010} or unfitted finite elements \cite{Belytschko2003_StructuredXFEM,Burman2015_CutFEM}, to avoid the complex issue of generating an exact hexahedral mesh of the domain. The needle, on the other hand, is modeled using Euler-Bernoulli beam elements having $6$ degrees of freedom ($3$ translations and $3$ rotations) per node.


\subsection{Error estimate and adaptive refinement}
\label{sec:SPR}

The accuracy of the outcome of surgical simulators depends on a number of factors. The latter can mainly be divided into two sources: modelling error and discretisation error. The first error source arises when we formulate a mathematical model for a physical problem, prompting the question: does the resulting model correctly describe the physical phenomenon? The second error source comes from the discretisation approach used by numerical methods such as the Finite Element (FE) method \cite{zienkiewicz1977finite}, or meshfree methods \cite{nguyen2008meshless,Griebel2003} to solve the mathematical model. In this paper, we assume the mathematical model to be correctly representing reality and focus only on the discretisation error.

By discretising the physical domain in order to solve numerically the mathematical model, the FE method introduces naturally a discretisation error in the solution. We estimate this error source by the superconvergent patch recovery (SPR) procedure~\cite{zienkiewicz1992superconvergent}. This information is then used to adaptively refine elements where the error is high.

The SPR is based on the following simple idea. The displacement field is obtained from solving the equilibrium equation, and the stress field is then computed by differentiating the polynomial approximation of the displacement field. In doing so, one looses numerical accuracy of the stress field (the differentiation reduces the polynomial degree by one). The stress field, obtained by the FEM, is possibly discontinuous across element boundaries. However, assuming that the original problem of the continuum has a smooth (continuous) solution, one can post-process the FEM solution to obtain higher accuracy. This post-processing is done on a collection of neighboring elements, which is called a patch.
Using SPR, the post-processed (smoothed) stress field $\bm{\sigma}^s$ is recovered from the stresses computed at the element centre. The idea of this technique is based on the fact that the stress and strain are more accurate at the superconvergent points (for the case of linear hexahedral elements used in this study, these points are at the centre point of elements), than anywhere else in the element. Using the least squares approach, the values of stress and strain at the superconvergent points are then employed to recover the nodal stress and strain.

The approximate error of an element $\Omega_{e}$ is defined as the energy norm
\begin{equation}
\label{eq:energyError}
\eta_{e} = \sqrt{\int_{\Omega^e} (\bm{\epsilon}^h - \bm{\epsilon}^s)^T (\bm{\sigma}^h - \bm{\sigma}^s)  \mathrm d \Omega },
\end{equation}
in which $\epsilon$ and $\sigma$ are strain and stress tensors, respectively. This error norm is indeed the difference between the FEM solution, denoted by superscript $h$, and an improved (recovered) solution, denoted by superscript $s$.
An element is marked for refinement if
\begin{equation}
\eta_{e} \ge \theta \eta_{M}, \; ~ 0< \theta <1, \; \text{where} ~ \eta_{M} = \max_{e} \eta_{e}.
\label{eq:max_rule}
\end{equation}

\begin{figure}[!htbp]
\centering
\includegraphics[width=1\columnwidth]{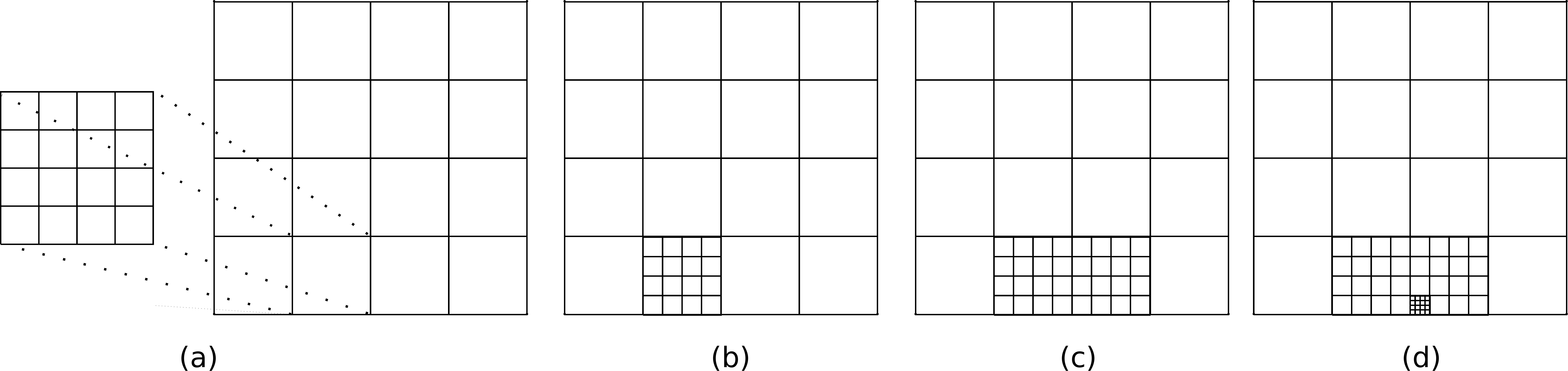}
\caption{Schematic presentation of template-based refinement: (a) A template for refinement can be defined prior to simulation; (b) first element is refined; (c) second element is refined; (d) the refinement can be nested.}
\label{fig:hexa_template}
\end{figure}

Based on a predefined template which is a set of nodes and an associated topology defining template elements, an element being marked for refinement is then replaced by the finer template elements using the mapped mesh method \cite{Grosland2009,Bui2016_TBME}.
%
%
The template nodes are added by using their natural coordinates (\ie{} the coordinates defined on the reference coordinate system with respect to the template).
The position $\mathbf{x}_j$, in Cartesian coordinates, of the template node $j$ is computed as
\begin{equation}
 \mathbf{x}_j = \sum_{i=1}^8 \mathbf{x}_i \xi_i^j, 
\end{equation}
where the summation is applied on the eight nodes $i$ of the removed hexahedron element.
The term $\xi_i^j$ in the interpolating polynomial (known as the shape function in the FEM context) can be seen as the \emph{relative} position of the node $j$ with respect to the node $i$.
%
A two dimensional schematic presentation of the refinement can be seen in \cref{fig:hexa_template}.
Note that the predefined template is not limited to regular elements, but can be as heterogeneous as desired (\ie{} the template with different element sizes). However, to have a good performance in FEM simulation, ill-shaped elements (\ie{} elements where one or more edges are much smaller than the largest) should be avoided in defining the template. Note also that we are not limited by the regularity of the mesh, and can start from any (reasonable) heterogeneous mesh as a starting point prior to refinement. Also, using this refinement technique, elements can be refined recursively, see \cref{fig:hexa_template}d.

\begin{figure}[!htbp]
\centering
\includegraphics[width=0.35\columnwidth]{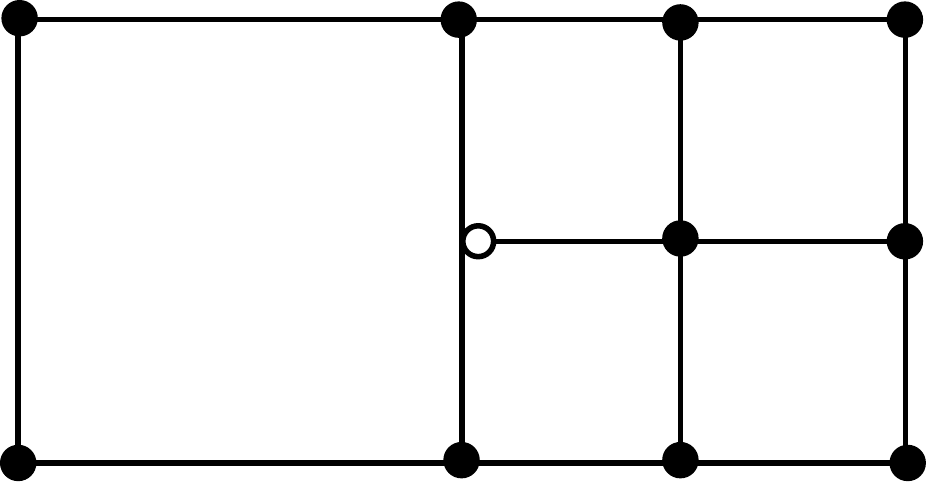}
\caption{Schematic representation of a hanging node (shown by the unfilled circle) in two dimensions.}
\label{fig:hanging_nodes}
\end{figure}

Since elements are refined according to the predefined templates, regardless of their neighbourhood, some nonconforming nodes (also known as hanging nodes) are generated. In FEM words, a nonconforming node is a node which is not shared by all elements around it, see \cref{fig:hanging_nodes}. These nodes are also called T-junctions since they form a T-shaped junction in the mesh. Solving the FEM problem without any treatment at T-junctions, displacement discontinuities at these nodes occur. To avoid that, Lagrange multipliers can be used to enforce the continuity at T-junctions, see \cref{sec:solveSystem}.

\subsection{Solving system equations with constraints}
\label{sec:solveSystem}
Based on \cref{eq:Ax=b}, the interaction between the needle (denoted by subscript $1$) and the tissue (denoted by subscript $2$) can be expressed by the following equation set
\begin{equation}
\label{eq:needle_tissue_interaction}
 \begin{pmatrix}
  \mathbf{A}_1 & \mathbf{0} & \mathbf{H}_1^T \\[0.3em]
  \mathbf{0} & \mathbf{A}_2 & \mathbf{H}_2^T \\[0.3em]
  \mathbf{H}_1 & \mathbf{H}_2 & \mathbf{0}
 \end{pmatrix} \begin{Bmatrix}
                d\mathbf{v}_1 \\[0.3em] d\mathbf{v}_2 \\[0.3em] \bm{\lambda}_i
               \end{Bmatrix}
               = \begin{Bmatrix}
                  \mathbf{b_1} \\[0.3em] \mathbf{b_2} \\[0.3em] \mathbf{0}
                 \end{Bmatrix},
\end{equation}
where $\bm{\lambda}_i$ is the Lagrange multiplier representing the \emph{interaction} between the needle and the tissue.
When refinement takes place in the tissue model, T-junctions (as explained above) can be handled by using constraints in the following form
\begin{equation}
\label{eq:Tjunctions}
 \begin{pmatrix}
  \mathbf{A}_2 & \mathbf{T}^T \\[0.3em]
  \mathbf{T} & \mathbf{0}
 \end{pmatrix}
 \begin{Bmatrix}
                d\mathbf{v}_2 \\[0.3em]
                \bm{\lambda}_t
\end{Bmatrix}
               = \begin{Bmatrix}
                  \mathbf{b_2} \\[0.3em]
                  \mathbf{0}
                 \end{Bmatrix}.
\end{equation}
where $\bm{\lambda}_t$ stands for Lagrange multipliers used for \emph{T-junctions}, and $\mathbf{T}$ expresses the dependence between the T-junction nodes (slave nodes) with respect to their parent nodes (master nodes).
We can see that \cref{eq:needle_tissue_interaction,eq:Tjunctions} have the same general form
\begin{equation}
\label{eq:generalConstraintEqs}
 \begin{pmatrix}
  \mathbf{A} & \mathbf{J}^T \\[0.3em]
  \mathbf{J} & \mathbf{0}
 \end{pmatrix}
 \begin{Bmatrix}
                \mathbf{x} \\[0.3em]
                \bm{\lambda}
\end{Bmatrix}
               = \begin{Bmatrix}
                  \mathbf{b} \\[0.3em]
                  \mathbf{0}
                 \end{Bmatrix}.
\end{equation}
\cref{eq:generalConstraintEqs} can be reformulated as
\begin{subequations}
 \begin{align}
  \mathbf{x} & = \underbrace{\mathbf{A}^{-1} \mathbf{b} }_{\mathbf{x}_{free}} - \mathbf{A}^{-1} \mathbf{J}^T \bm{\lambda}, \label{eq:x} \\
  \mathbf{J} \mathbf{A}^{-1} \mathbf{J}^T \bm{\lambda} & =  \mathbf{J} \underbrace{\mathbf{A}^{-1} \mathbf{b} }_{\mathbf{x}_{free}}, \label{eq:lambda}
 \end{align}
\end{subequations}
in which, $\mathbf{x}_{free}$ can be seen as the solution of the unconstrained system $\mathbf{A} \mathbf{x} = \mathbf{b}$. Therefore, \cref{eq:generalConstraintEqs} can be solved in three steps as
\begin{enumerate}
\setlength{\itemindent}{1em}
 \item[Step 1.] ~Factorize the matrix $\mathbf{A}$ to have its inverse $\mathbf{A}^{-1}$, and solve for $\mathbf{x}_{free}$,
 \item[Step 2.] ~Solve Lagrange multipliers $\bm{\lambda}$ from \cref{eq:lambda},
 \item[Step 3.]~Once $\bm{\lambda}$ is available, $\mathbf{x}$ can be obtained from \cref{eq:x} by using $\mathbf{x}_{free}$.
\end{enumerate}

\section{Results}
\label{sec:Results}
In this section, we present several numerical studies in order to demonstrate the efficiency of our method. To show the benefits of employing adaptive refinement controlled by error estimates, we first study the simulation of needle insertion into a phantom tissue of simple geometry. Then, we study a more complex simulation of deep-brain stimulation (DBS) lead insertion. In DBS, a cannula is inserted into the brain tissue in order to reach an STN target. Leaving an electrode inside the brain, the cannula is then pulled back until completely outside of the brain. By these studies, we provide insight into the mechanical behaviour of the brain tissue in response to the lead insertion, and analyse the effect of mesh adaptivity on the solution.

DBS is an effective approach to alleviate the symptoms of neuro-degenerative diseases such as Parkinson's disease. During most DBS interventions, the craniotomy leads to a shift of the brain within the skull, because of the cerebrospinal fluid which is drained out of the skull cavity and ceases to provide buoyancy to the brain. Once the electrode has been inserted, the brain regains its original position.

One of the difficulties during DBS interventions is that the electrode displaces as the brain comes back to its original position, before brain shift. Therefore, it is possible that the electrode tip displaces, so that the electrode does not stimulate the proper target anymore, which requires a new operation.

Simulations of the type that we are presenting here have the potential to help surgeons prepare for interventions by investigating where the electrode tip should be located, so that it stimulates the target \emph{after the brain shifts back to the configuration it had prior to craniotomy-induced brain shift}.

\subsection{Impact of local mesh refinement on displacement field}

\begin{figure}[!htbp]
\centering
 \includegraphics[width=0.5\columnwidth]{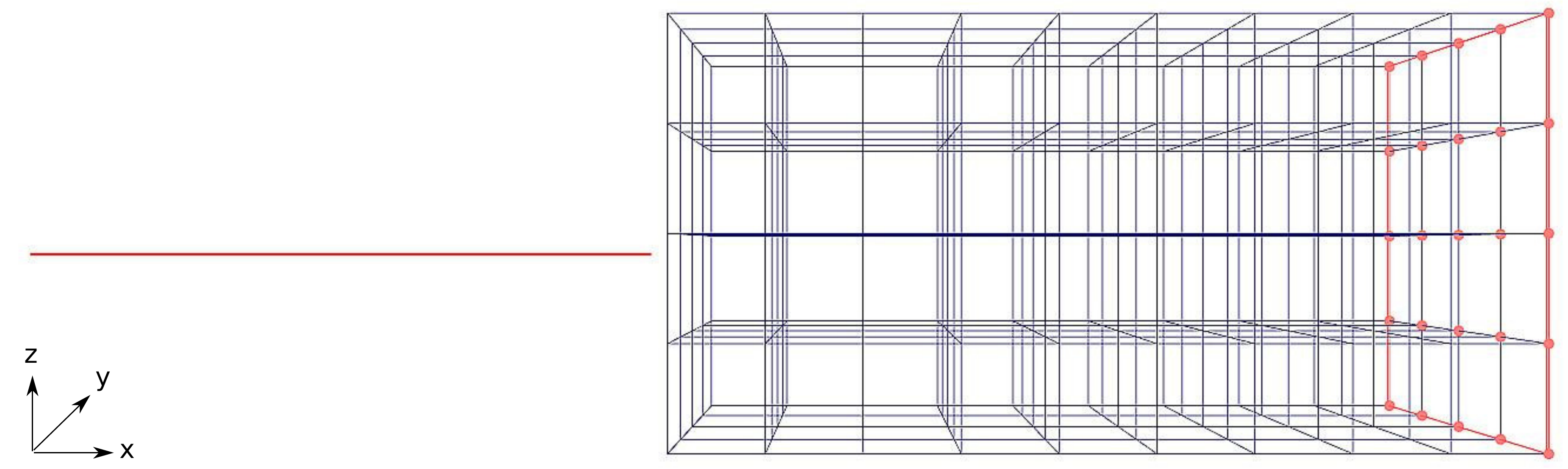}
\caption{Schematic illustration of the needle insertion problem into the tissue with a simple geometry being clamped at the right end surface as the boundary condition.}
\label{fig:needle_insertion_beam_sharpen}
\end{figure}

The effect of adaptive refinement on the compromise between the computational time and the solution accuracy is studied through a needle insertion simulation into a phantom tissue. In this study, the mechanical properties are taken as follows. The Young's modulus is set to $50$~MPa for the needle, and $10$~MPa for the tissue, whereas $0.4$ and $0.3$ are set for the Poisson's ratio of the tissue and the needle, respectively.
%
%
It is noted that, for this heuristic study, these parameters are chosen arbitrarily (but are physically meaningful for a needle-tissue interaction problem). \cref{fig:needle_insertion_beam_sharpen} shows an illustration of the needle insertion problem.

As mentioned above, we employ a linear elastic model based on corotational formulation to model the behaviour of the needle as well as the tissue. The tissue geometry has the dimension of $4 \times 2 \times 2$~cm. The length of the needle and its radius are of $3.2$~cm and $0.1$~cm, respectively.

\begin{figure}[!htbp]
\centering
\includegraphics[width=1\columnwidth]{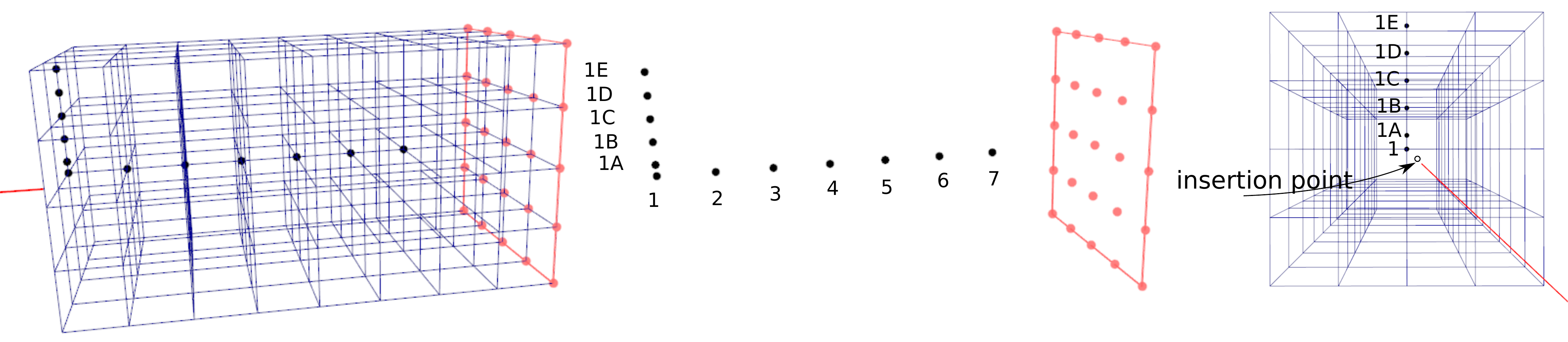}
\caption{The displacement during insertion is measured at the predefined points, namely, $1$, $2$, $3$, $4$, $5$, $6$, $7$ which are distributed in the direction of the needle shaft, and $1A$, $1B$, $1C$, $1D$, $1E$ that lie in the direction normal to the needle shaft.}
\label{fig:points_definition}
\end{figure}

The mesh is subsequently refined in a uniform manner ($8\times 4\times 4$, $16\times 8\times 8$, $32\times 16\times 16$ elements), as well as in an adaptive manner. By the adaptive refinement scheme, we start simulations with the coarse mesh ($8 \times 4 \times 4$ nodes) and adaptively refine the mesh during the simulation thanks to the error estimate, see~\cref{sec:Method}.
During needle insertion, we measure the displacement of the points defined in~\cref{fig:points_definition}.

%

\begin{figure}[!htbp]
    \begin{minipage}[b]{.5\linewidth}
          \centering
          \includegraphics[width=1\columnwidth]{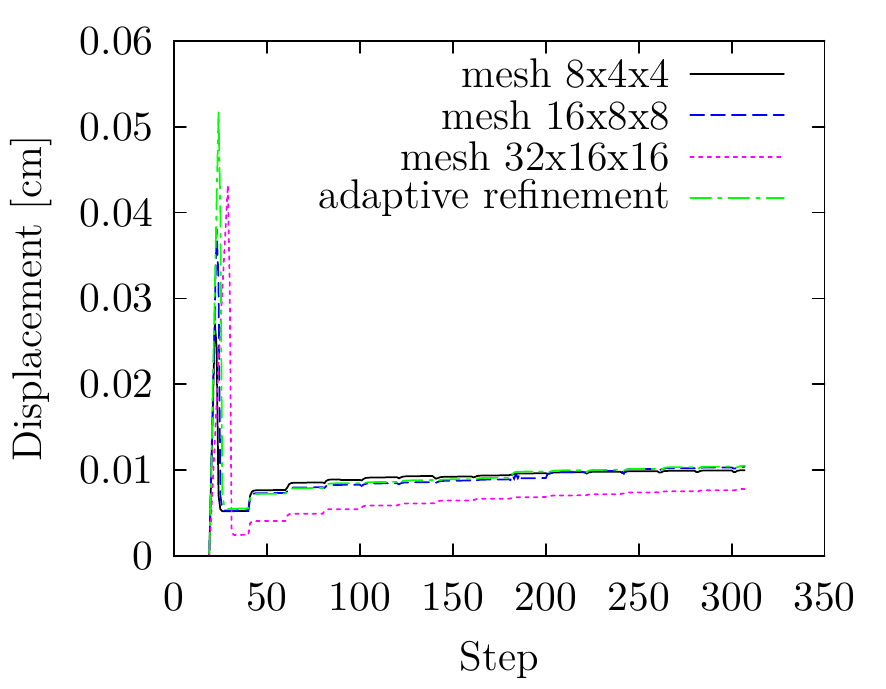}
          \subcaption{Point $1$.}
          \label{fig:displacement_radial_point_1_vary_mesh}
    \end{minipage}
    ~
    \begin{minipage}[b]{.5\linewidth}
          \centering
          \includegraphics[width=1\columnwidth]{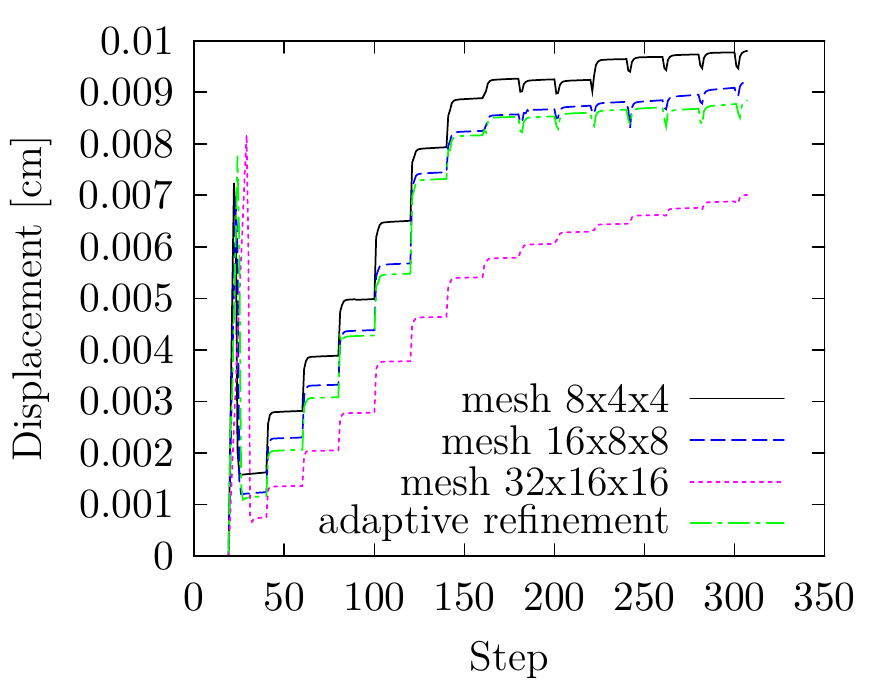}
          \subcaption{Point $3$.}
          \label{fig:displacement_along_needle_point_3_vary_mesh}
    \end{minipage}

    \begin{minipage}[b]{.5\linewidth}
          \centering
          \includegraphics[width=1\columnwidth]{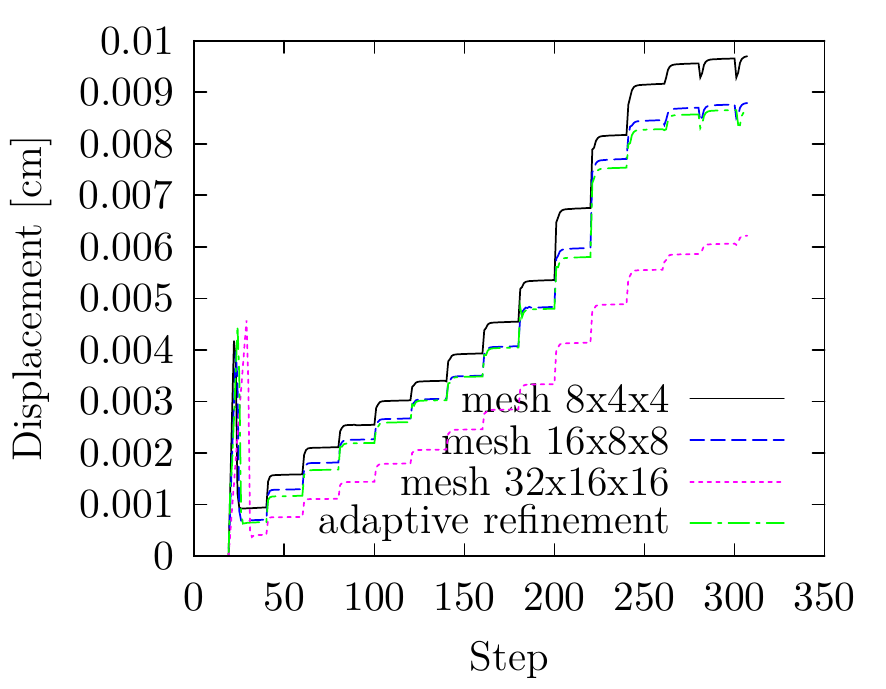}
          \subcaption{Point $5$.}
          \label{fig:displacement_along_needle_point_5_vary_mesh}
    \end{minipage}
    ~
    \begin{minipage}[b]{.5\linewidth}
          \centering
          \includegraphics[width=1\columnwidth]{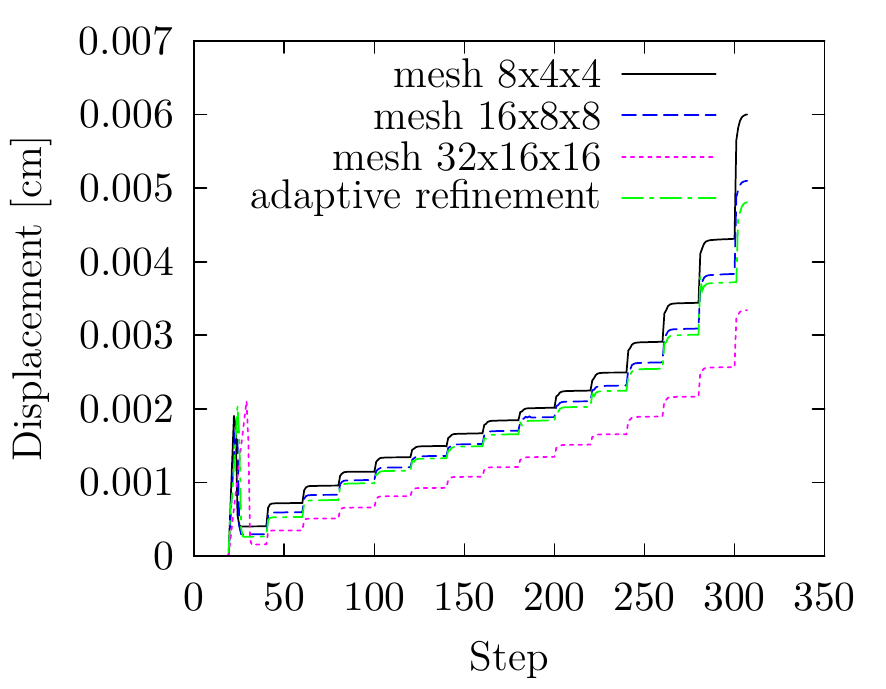}
          \subcaption{Point $7$.}
          \label{fig:displacement_along_needle_point_7_vary_mesh}
    \end{minipage}
\caption{During needle insertion, the displacement of the points located in the direction of the needle shaft, uniform and adaptive refinement schemes. The numbers of degrees of freedom for the meshes listed in the graphs (in the label order) are $675$, $4\,131$, $28\,611$, and $1\,461$ (for adaptive refinement scheme it is the maximum at the end of insertion step). It is seen that the displacement solution of adaptive refinement scheme agrees well with that of fine mesh when certain criterion of adaptivity is appropriately chosen (here $\theta=0.3$). Moreover, it is observed that the finer the mesh is used, the smaller the displacements at those points are.}\label{fig:displacement_along_needle}
\end{figure} 

\begin{figure}[!htbp]
    \begin{minipage}[b]{.5\linewidth}
              \centering
              \includegraphics[width=1\columnwidth]{displacement_radial_point_1_vary_mesh.pdf}
              \subcaption{Point $1$.}
              \label{fig:displacement_radial_point_1_vary_mesh_1}
    \end{minipage}
    ~
    \begin{minipage}[b]{.5\linewidth}
              \centering
              \includegraphics[width=1\columnwidth]{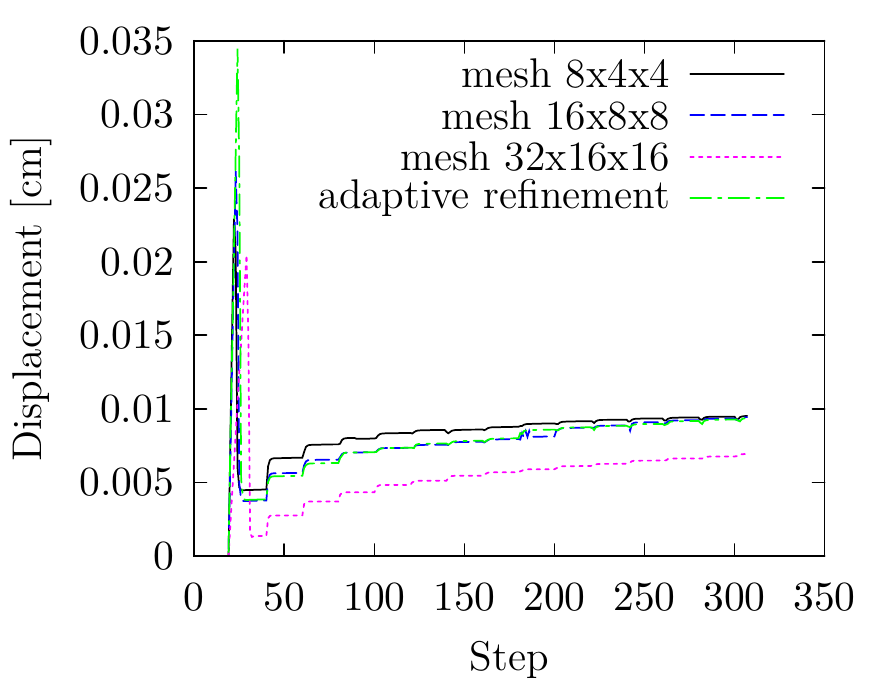}
              \subcaption{Point $1A$.}
              \label{fig:displacement_radial_point_1A_vary_mesh}
    \end{minipage}

        \begin{minipage}[b]{.5\linewidth}
              \centering
              \includegraphics[width=1\columnwidth]{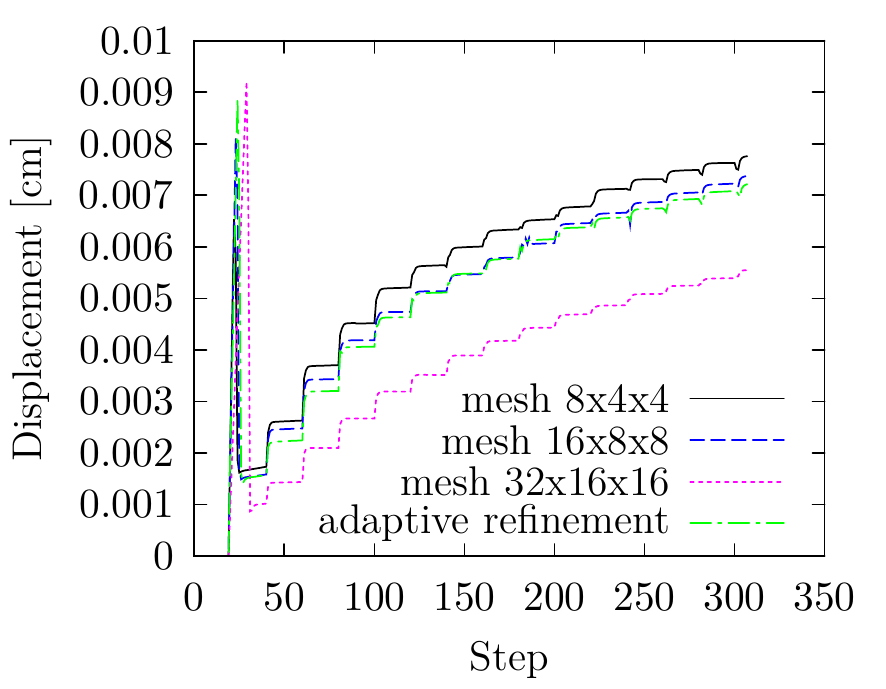}
              \subcaption{Point $1C$.}
              \label{fig:displacement_radial_point_1C_vary_mesh}
    \end{minipage}
    ~
        \begin{minipage}[b]{.5\linewidth}
              \centering
              \includegraphics[width=1\columnwidth]{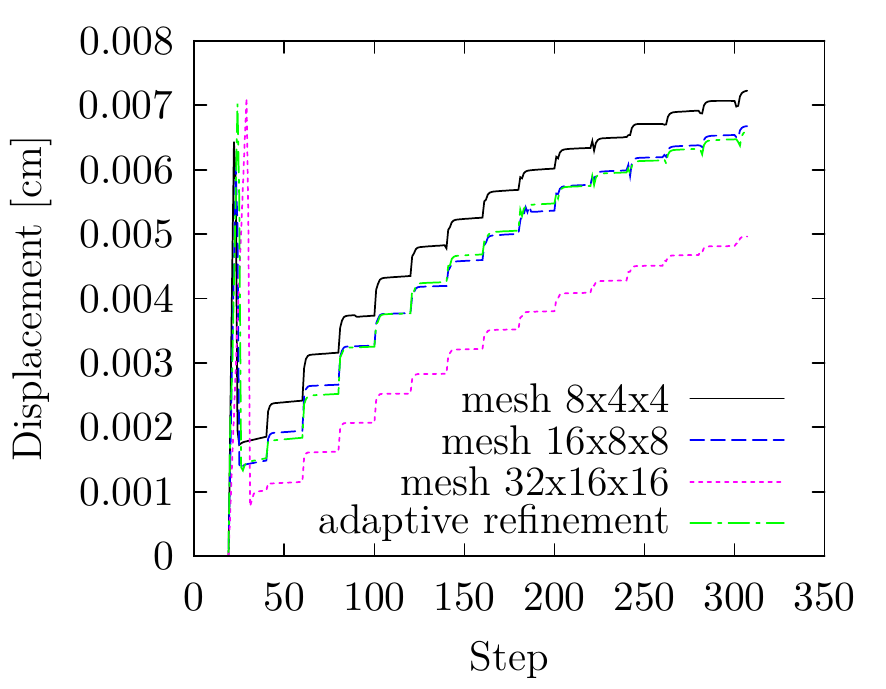}
              \subcaption{Point $1E$.}
              \label{fig:displacement_radial_point_1E_vary_mesh}
    \end{minipage}
\caption{The displacement of the points located in the radial direction, uniform and adaptive refinement schemes, during needle insertion. The numbers of degrees of freedom for the meshes listed in the graphs (in the label order) are $675$, $4\,131$, $28\,611$, and $1\,461$ (for adaptive refinement scheme it is the maximum at the end of insertion step). The same observations can be made as in~\cref{fig:displacement_along_needle}.}\label{fig:displacement_radial}
\end{figure}

Note that the displacements of all points is not shown here. \cref{fig:displacement_along_needle} shows the displacement of some points distributed in the direction of the needle shaft, and \cref{fig:displacement_radial} shows the displacement of some points distributed in the direction normal to the needle shaft (\ie{} radial).
These results reveal two distinct features. Firstly, the displacement solution of the adaptive refinement scheme agrees well with that of fine mesh when the  adaptivity criterion is appropriately chosen (here $\theta=0.3$). Secondly, it is observed that after puncture, the finer the mesh is used, the smaller the displacements at those points are. This observation is explained by two phenomena. First, a finer mesh produces softer behaviour for the tissue, i.e. the displacement does not ``propagate'' as easily from the needle shaft to the point where we measure the displacement (refer to \cite{Bui2016_TBME} for details). Second, needle insertion leads to localisation of displacements, strains and stresses, due to the small size of the needle compared to the organ.

\begin{figure}[!htbp]
\centering
    \begin{minipage}[b]{.5\linewidth}
      \centering\includegraphics[width=1\columnwidth]{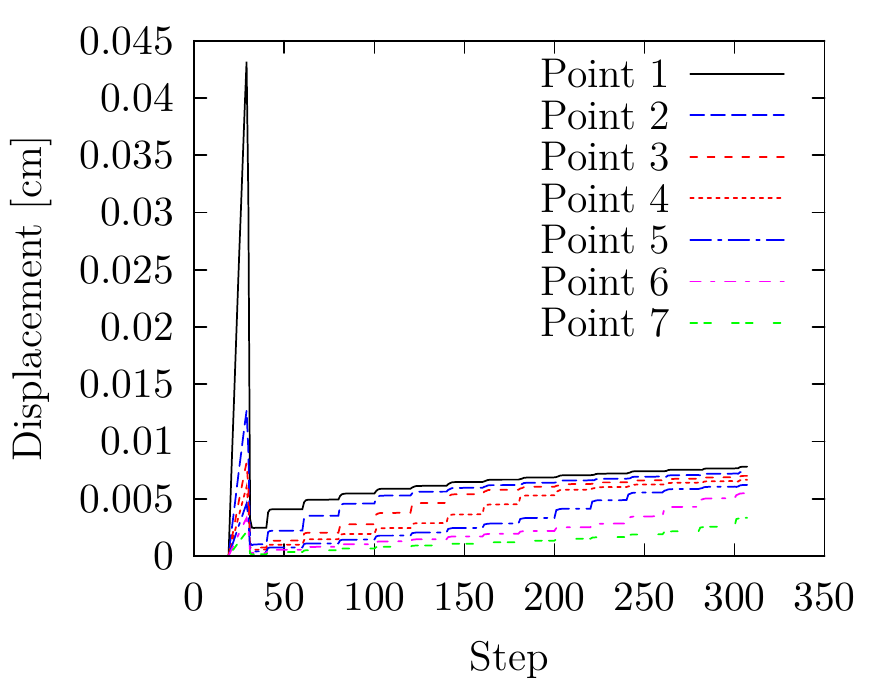}
      \subcaption{In needle direction.}
      \label{fig:displacement_allPoints_mesh32x16x16}
    \end{minipage}%
    \begin{minipage}[b]{.5\linewidth}
        \centering\includegraphics[width=1\columnwidth]{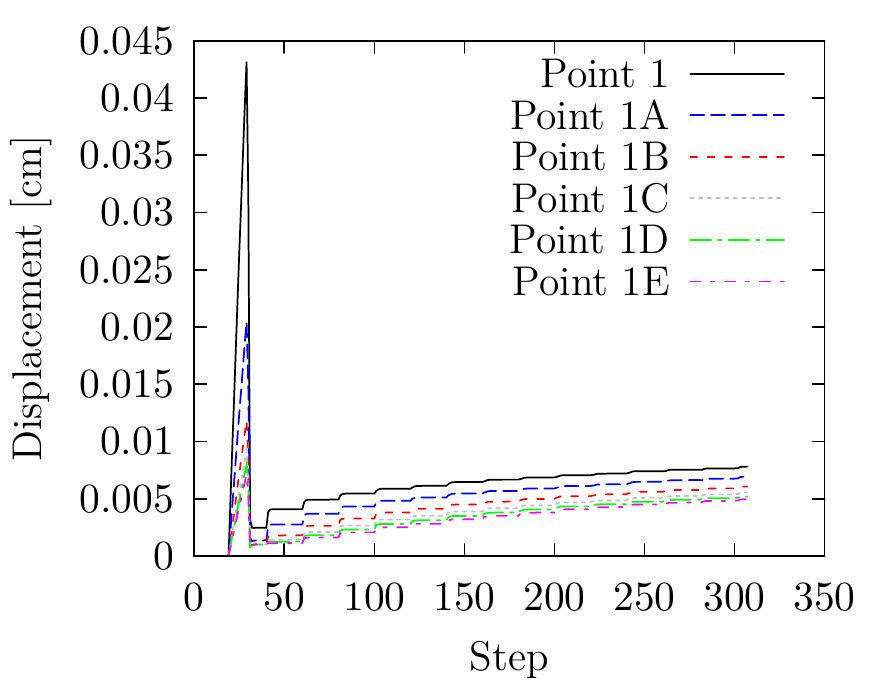} 
        \subcaption{In radial direction.}
        \label{fig:displacement_radial_point_1_1A_1B_1C_1D_1E_mesh32x16x16}
    \end{minipage}
\caption{Displacement of the points during needle insertion, when the  $32\times 16\times 16$ mesh is employed. The further the point from the tissue surface, or from the needle shaft, is found, the smaller the displacement of that point is.}\label{fig:displacement_radial_along_mesh32x16x16}
\end{figure}

\cref{fig:displacement_radial_along_mesh32x16x16} shows the displacement at all predefined points during insertion when  mesh $32\times 16\times 16$ is employed. It is seen that the further the point from the tissue surface is (\cref{fig:displacement_allPoints_mesh32x16x16}), or the further the point from the needle shaft is (\cref{fig:displacement_radial_point_1_1A_1B_1C_1D_1E_mesh32x16x16}), the smaller the displacement of that point is. The physical relevance of this observation is explained in the previous paragraph.

\subsection{Electrode implantation in DBS}

We study the simulation of DBS lead placement by inserting a cannula into a brain model until it reaches the predefined STN target. Then, the cannula is retracted while keeping an electrode inside the brain. As in \cite{Bilger2011}, the Young's modulus of $10$ GPa is set for the cannula and for the electrode, while the Young's modulus of the brain tissue is set to $6$ kPa. The Poisson's ratio is set as $0.45$ for the brain tissue, and $0.3$ for both the cannula and the electrode. The radii of the cross sections are set to $3$ mm and $0.7$ mm for the cannula and electrode, respectively. The frictional coefficient between the cannula and the brain tissue is established at $0.05$. The adaptivity parameter $\theta$ is set to 0.6. From our studies, we found that, with minimum DOFs, this choice of the adaptivity parameter gives the best results for DBS simulation of electrode placement. Since the brain tissue is very soft, we set the penetration strength at the brain surface and the cutting strength to $0.01$ N.

It is worth mentioning that, in this study, simple boundary conditions around the brain tissue are taken into account. Indeed, we simply consider clamped bilateral constraints around the area of the optic nerves and the brainstem. We believe that this choice of boundary conditions (for the sake of simplicity) does not affect the efficiency of the adaptive refinement algorithm. However, more complex boundary conditions for the brain with respect to the skull can be found in \cite{Wittek2007c,Bilger2011}. We are currently investigating the effects of the choice of boundary conditions on the results for this particular case. In general, for brain shift problems, it is now known that the type of boundary conditions does not have a strong effect on the resulted brain shift, see, \eg{}, \cite{Wittek2005,Joldes2009}. However, systematic studies are still missing from the literature for the problem of DBS simulation.

Two strategies of mesh refinement are employed. The uniform refinement consists of refining each hexahedron of the mesh into 8 smaller hexahedra. In the second strategy, called adaptive refinement, we only refine the elements that fulfil the adaptivity condition~\cref{eq:max_rule}.

\begin{figure}[!htbp]
\centering
\begin{minipage}[b]{.3\linewidth}
\centering
\includegraphics[width=1\columnwidth]{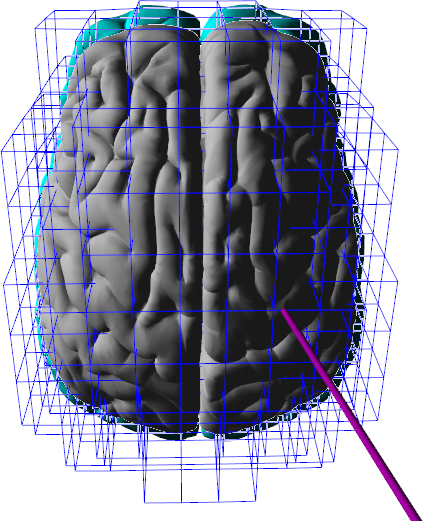}\\ (a)
\end{minipage}
 \qquad 
\begin{minipage}[b]{.38\linewidth}
\centering
\includegraphics[width=1\columnwidth]{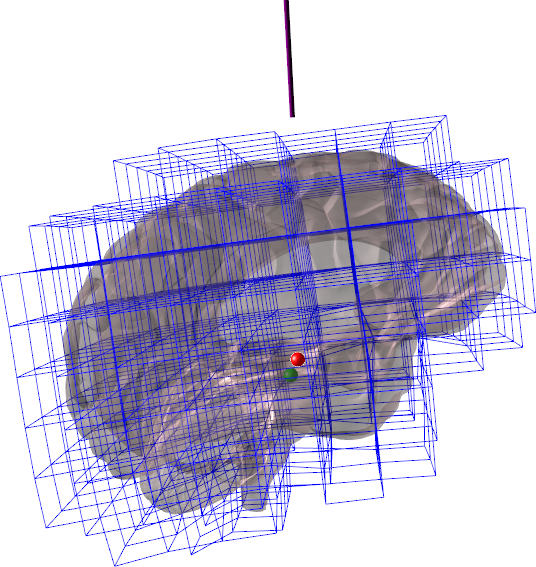}\\ (b)
\end{minipage}
\caption{Simulation of brain shift prior to insertion of the cannula for electrode implantation. Initial position of the brain is shown by cyan colour (a). The position of the STN target after brain shift is shown by the green sphere, while its initial position is shown by the red one (b). The displacement of the STN target due to brain shift is about $6.5$ mm. }\label{fig:show_brain_shift}
\end{figure}

\begin{figure}[!htbp]
\centering
\includegraphics[width=0.65\textwidth]{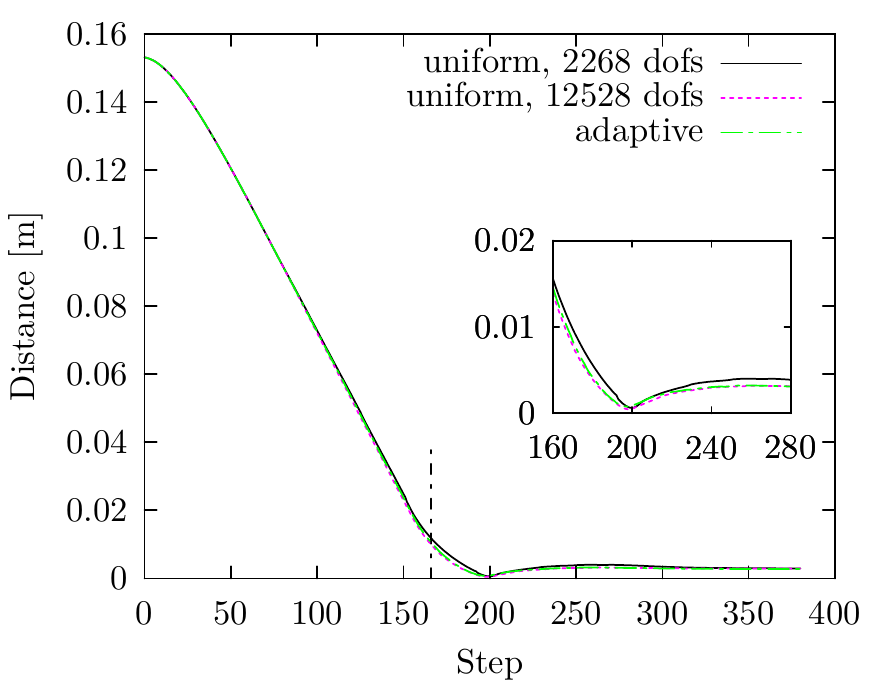}
\caption{The distance from the electrode tip to the STN target. An STN target is fixed inside the brain. The cannula is inserted into the brain so that it can reach the STN target. Then the electrode is left while retracting the cannula. The vertical line at step $166$ indicates the moment when the cannula has the trend to be pulled back. It is seen that using an adaptive refinement scheme, the distance obtained between the electrode tip and the STN target is as good as using the fine uniform mesh. However, a significant gain in computational time is observed. The maximum number of degrees of freedom for the adaptive refinement is 3,135. For a similar solution profile, compared to a uniformly refined mesh, this represents a reduction in the problem size of approximately $4$.}\label{fig:distance}
\end{figure}

In~\cref{fig:show_brain_shift}, brain shift prior to insertion is shown. ~\cref{fig:distance} shows the distance between the electrode tip and the STN target during insertion and retraction of the cannula. The vertical line in~\cref{fig:distance} indicates the point in time when the cannula has the trend to be pulled back. It is observed that from that time, the distance between the electrode tip and the STN target still decreases. This is justified by the fact that the brain tissue is unloaded from insertion forces and displaces toward the direction from which the needle was inserted. The electrode, on the other hand keeps moving forward due to inertial forces. When the cannula is being retracted, the brain tissue deforms and moves backward, and thus increases the electrode-STN target distance accordingly. When the cannula is completely outside the brain tissue, that distance is stabilised at around $2.9$ mm.

Moreover, \cref{fig:distance} reveals that the distance from the electrode tip to the STN target obtained by using our adaptive refinement scheme agrees well with that when a fine uniform mesh is employed. However, an important gain in computational time can be observed, since the maximum number of DOFs for the adaptive scenario is only $3\,135$ as compared to $12\,528$ DOFs of the fine uniform mesh. This reduction in the problem size is approximately of the order of $4$. 

\begin{figure}[!htbp]
\centering
\includegraphics[width=0.65\textwidth]{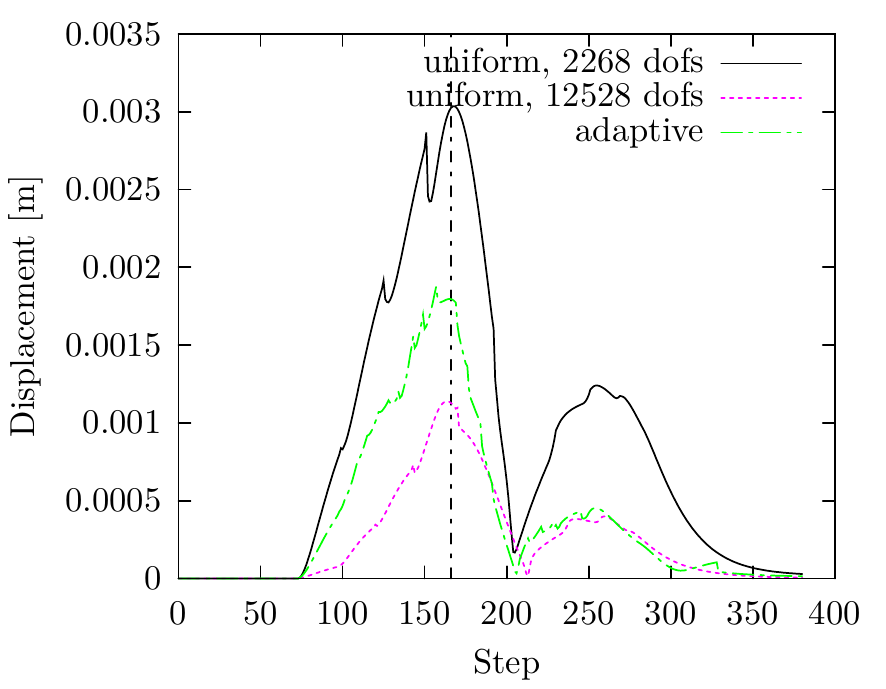}
\caption{Displacement of STN target (with respect to its position after brain shift) during cannula insertion and retraction (the same simulations as in \cref{fig:distance}). The vertical line, same as in \cref{fig:distance}, shows the stage when the cannula is going to be removed. The maximum number of DOFs for the adaptive refinement is 3135.}\label{fig:displacement_parkinson_target}
\end{figure}

\cref{fig:displacement_parkinson_target} shows the displacement of the STN target (with respect to its position after brain shift) when inserting the cannula, leaving the electrode, and retracting the cannula. It can be seen that at around the step $75$, the cannula starts penetrating into the brain tissue, and thus provokes the displacement of the STN target inside the brain. This displacement increases along with the insertion of the cannula and the electrode into the brain, due to cutting of the brain tissue and friction on the cannula shaft. At the step $166$, when the cannula starts undergoing retraction, the brain tissue is then unloaded, and this causes the STN target to move backward. This phenomenon is seen by the decrease of the STN target displacement up to the step between $200$ and around $225$ (depending on the mesh employed, refer to \cref{fig:displacement_parkinson_target}). After that, since the cannula is bearing retraction, and because of friction, the brain tissue follows the cannula in the direction of the retraction. As a consequence, the displacement of the STN target increases again. The closer to the brain surface the cannula tip is, the smaller the magnitude of frictional force along the cannula shaft is, and therefore, at some stage, the STN target displacement decreases again. And, that displacement continuously decreases when the cannula is fully retracted from the brain tissue, as observed at the final stage of \cref{fig:displacement_parkinson_target}.

\begin{figure}[!htbp]
\centering
    \begin{minipage}[b]{.3\linewidth}
              \centering
              \includegraphics[width=1\columnwidth]{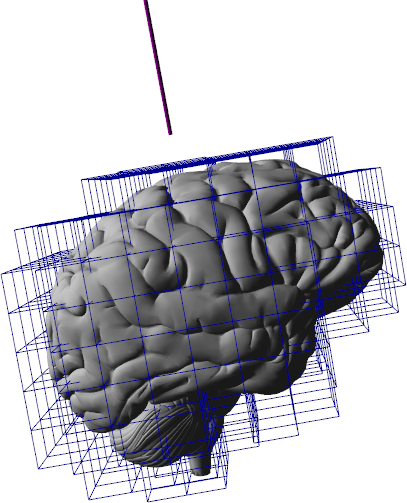}
              \subcaption{}
              \label{fig:init_simulation_cropped}
    \end{minipage}
    \qquad 
    \begin{minipage}[b]{.3\linewidth}
              \centering
              \includegraphics[width=1\columnwidth]{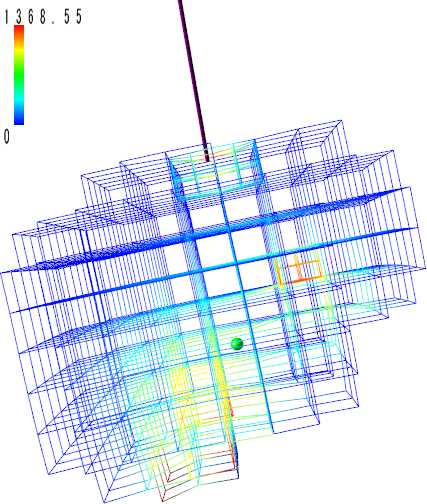}
              \subcaption{}
              \label{fig:refinement_at_penetration_cropped}
    \end{minipage}
      
    \begin{minipage}[b]{.3\linewidth}
              \centering
              \includegraphics[width=1\columnwidth]{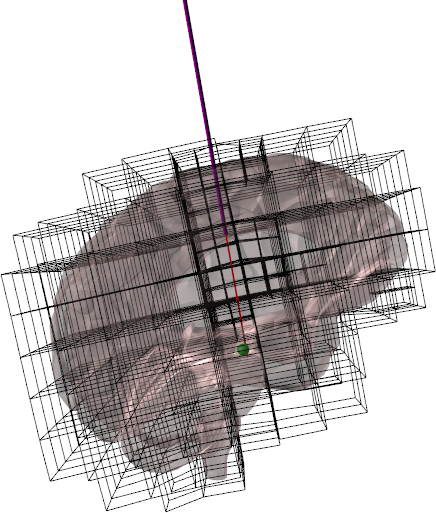}
              \subcaption{}
              \label{fig:cannula_inserted_cropped}
    \end{minipage}
    \qquad 
    \begin{minipage}[b]{.3\linewidth}
              \centering
              \includegraphics[width=1\columnwidth]{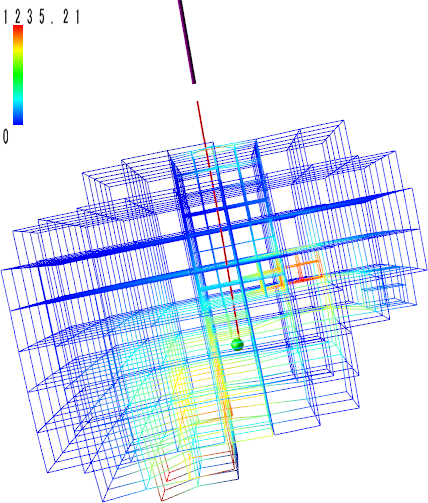}
              \subcaption{}
              \label{fig:cannula_retracted_cropped}
    \end{minipage}
\caption{Visualisation of the electrode implantation simulation of DBS (\subref{fig:init_simulation_cropped}). Brain shift occurs prior to cannula insertion. The STN target during the simulation is shown at different stages by a green sphere in (b,c,d). When the cannula tip is in contact with the brain tissue surface, it leads to element refinement guided by error estimate, shown in (\subref{fig:refinement_at_penetration_cropped}), coloured by Von Mises stress magnitude. When the cannula has reached the STN target, it undergoes a retraction. The cannula is being retracted while the electrode is left inside the brain, shown in (\subref{fig:cannula_inserted_cropped}). The cannula is completely retracted, adaptive refinement has occurred along the trajectory of the cannula, shown in (\subref{fig:cannula_retracted_cropped}).}
\label{fig:brain_patterns}
\end{figure}

Visualisation of the simulation of DBS lead implantation during insertion and retraction at some stages can be seen in~\cref{fig:brain_patterns}.
\cref{fig:displacement_parkinson_target} reveals that, during the insertion phase, the result of the adaptive mesh does not closely agree with that of the uniform fine mesh. This is evidenced by the fact that the refinement has begun but has not reached the STN target yet, as can be partly seen in \cref{fig:refinement_at_penetration_cropped}. However, from the stage when the cannula reaches the STN target, to the end of the simulation when it is completely pulled out of the brain tissue, the result of the adaptive mesh agrees well with that of the uniform fine mesh, thanks to the refinement occurring along the cannula trajectory, as seen in~\cref{fig:cannula_retracted_cropped}.

It is again observed that the use of the proposed adaptive refinement scheme is very relevant because it guarantees the solution accuracy close to the fine meshing schemes, with computational time suitable for real-time simulation.

\section{Conclusions}
\label{sec:Conclusions}

We have presented a structured study to answer the essential, but rarely addressed, question of accuracy in surgical simulations. The original contribution of our paper is the use of an a posteriori estimate of the discretisation error to automatically drive local adaptive mesh refinement during needle insertion.

From the study on the displacement solution of two different problems (needle insertion into a simple phantom geometry with stiff mechanical properties, and the simulation of electrode implantation for deep brain stimulation (DBS) with softer mechanical properties for the brain), two major conclusions can be drawn:
\begin{itemize}
\item The outcome of the proposed adaptive refinement scheme does not depend on the mechanical parameters of the tissue, such as Young's modulus and Poisson's ratio.
\item With a suitably chosen refinement criterion, the adaptive solution agrees well with the solution on a fine uniform mesh, while saving computational time to make it feasible for real-time simulations. The computational time savings in the particular case we tackled are of the order of $10$ (more details can be found in \cite{Bui2016_TBME}).
\end{itemize}

Our work is limited in several ways which we are currently investigating:
\begin{itemize}
\item We tackled only the discretisation error, and avoided the difficult problem of ``model error''. For example, we assumed the brain to behave in a corotational manner. Although this is corroborated by the literature,  we do not have physical evidence that this assumption is true in general, nor as to when it may break down. 
\item We assumed simple boundary conditions around the brain. This should be investigated in more detail to assess the effect of both geometrical and boundary condition uncertainty on the outcome of the simulations.
\item Soft tissue properties vary from patient to patient by up to a few orders of magnitude. We have been working on robust and systematic approaches to quantifying these uncertainties \cite{Hauseux2017}. We will use these methods to quantify the effects of variability in the material parameters on the motion of the target and of the electrode during DBS.
\end{itemize}

%

\section*{Acknowledgments}

St\'ephane Bordas, Satyendra Tomar and Huu Phuoc Bui thank partial funding for their time provided by the European Research Council Starting Independent Research Grant (ERC Stg grant agreement No. 279578) RealTCut ``Towards real time multiscale simulation of cutting in non-linear materials with applications to surgical simulation and computer guided surgery''. We also also grateful for the funding from the Luxembourg National Research Fund (INTER/MOBILITY/14/8813215/CBM/Bordas and INTER/FWO/15/10318764).

St\'ephane Cotin thanks for funding of European project RASimAs (FP7 ICT-2013.5.2 No. 610425).

The funding from the University of Strasbourg Institute for Advanced Study (BPC 14/Arc 10138) for the first author is gratefully acknowledged.

Huu Phuoc Bui thanks Dr. Davide Baroli and Mr. Vahid Shahrabi for their useful helps about PyMesh (\url{https://github.com/qnzhou/PyMesh}) and Meshmixer (\url{http://www.meshmixer.com}).

The authors would also like to thank Dr. Alexandre Bilger and Mr. R\'emi Bessard Duparc for their general helps.

\bibliographystyle{wileyj}
\bibliography{refs}

\end{document}